\documentclass[aps,prd,preprint,floatfix,nofootinbib,superscriptaddress,showpacs]{revtex4}
 \usepackage{graphicx}
 \def\deg{^\circ}
\def\deg{^\circ}

\def\Z0{${\em Z^0\/}$}

\def\r#1 {$^{#1}$}

\newcommand{\gevc} { {\rm GeV/c}}
\newcommand{\gevcc}{ {\rm GeV/c^2}}

\def\gepsfcentered#1{
  \def\testit{#1}
  \def\lbracket{[}
  \ifx\testit\lbracket
    \let\dofilecmd=\gepsfwithopt
  \else
    \let\dofilecmd=\gepsfnoopt
  \fi
  \dofilecmd}

\def\gepsfnoopt#1{
  \begin{center}
  \leavevmode
  \epsffile{#1}
  \end{center}}

\def\gepsfwithopt#1 #2 #3 #4]#5{
  \begin{center}
  \leavevmode
  \gepsfmaxx=0.94\textwidth
  \epsffile[#1 #2 #3 #4]{#5}
  \end{center}}

\newdimen\gepsfmaxx
\def\epsfsize#1#2{
  \ifnum \epsfxsize=0
    \ifnum \epsfysize=0
      \ifnum #1 > \gepsfmaxx
        \gepsfmaxx
      \else
        #1
      \fi
    \else
      \epsfxsize
    \fi
  \else
    \epsfxsize
  \fi
}

\begin{document}

 \bibliographystyle{apsrev}
%=======
% Title
%=======
 \title {Study of sequential semileptonic decays
         of $b$ hadrons produced at the Tevatron}
%==========
% Authors (First the desired ordering of institutions, then the authors)
%==========
 \affiliation{Laboratori Nazionali di Frascati, Istituto Nazionale 
         di Fisica Nucleare, Frascati, Italy\\}
 \affiliation{Fermi National Accelerator Laboratory, Batavia, 
         Illinois 60510, USA \\}
 \affiliation{Istituto Nazionale di Fisica Nucleare, University and
         Scuola Normale Superiore of Pisa, I-56100 Pisa, Italy\\}
 \author{G.~Apollinari}
 \affiliation{Fermi National Accelerator Laboratory, Batavia, 
         Illinois 60510, USA \\}
 \author{M.~Barone}
 \affiliation{Laboratori Nazionali di Frascati, Istituto Nazionale 
         di Fisica Nucleare, Frascati, Italy\\}
 \author{I.~Fiori}
 \affiliation{Istituto Nazionale di Fisica Nucleare, University and
         Scuola Normale Superiore of Pisa, I-56100 Pisa, Italy\\}
 \author{P.~Giromini}
 \author{F.~Happacher}
 \author{S.~Miscetti}
 \author{A.~Parri}
 \author{F.~Ptohos}
 \altaffiliation[Present address:]{ University of Cyprus, 1678 Nicosia, 
	Cyprus\\}
 \affiliation{Laboratori Nazionali di Frascati, Istituto Nazionale 
         di Fisica Nucleare, Frascati, Italy\\}
%%%%%%%%%%%%%%%%%%%%%
 \begin{abstract}
   We present a study of rates and kinematical properties of lepton pairs 
   contained in central jets with transverse energy $E_T \geq 15$ GeV that
   are produced at the Fermilab Tevatron collider. We compare the data to
   a QCD prediction based on the {\sc herwig} and {\sc qq} Monte Carlo
   generator programs. We find that the data are poorly described by the
   simulation, in which sequential semileptonic decays of single $b$ quarks 
   ($b \rightarrow l\; c \; X$ with $c \rightarrow l\; s\;  X$) are the
   major source of such lepton pairs. \\
 \end{abstract}  
%==== 
% PACS numbers and preprint number with the macros
%====
 \preprint{FERMILAB-PUB-05-271-E} 
%======
% End of front page
%======
 \maketitle
%%%%%%%%%%%%%%%%%%%%%
%
%%%%%%%%%%%%%%%%%%%%%%%%%%%%%%%%%%%%%%%%%%%%%%%%%%
 \section {Introduction}  \label{sec:ss-intro}
%%%%%%%%%%%%%%%%%%%%%%%%%%%%%%%%%%%%%%%%%%%%%%%%%%
   This study of sequential semileptonic decays of $b$ hadrons completes the
   review of the heavy flavor properties of jets produced at the Fermilab 
   Tevatron collider presented in Ref.~\cite{apo}. The data set, collected
   with the Collider Detector at Fermilab (CDF) in the $1992 - 1995$ collider
   run, consists of events with two
   or more jets with transverse energy $E_T \geq 15$ GeV and pseudorapidity
   $|\eta| \leq 1.5$, and is the same as that used in Ref.~\cite{apo}.
   The heavy flavor purity of the sample is enriched by requiring that at 
   least one of the jets contains a lepton ($e$ or $\mu$) with transverse 
   momentum larger than $8 \; \gevc$. The jet containing the lepton is 
   referred to as lepton-jet, whereas the jets recoiling against the 
   lepton-jet  are called away-jets. Since these events have been acquired 
   by triggering on the presence of a lepton with $p_T \geq 8\; \gevc$, 
   we call electron and muon data the samples with an  electron- and
   muon-jet, respectively. Jets containing hadrons with heavy flavor are 
   identified using the CDF silicon micro-vertex detector (SVX) to locate
   secondary vertices produced by the decay of $b$ and $c$ hadrons inside 
   a jet. These vertices ({\sc secvtx} tags) are separated from the primary
   vertex as a result of the long $b$ and $c$ lifetime. The $b$- and 
   $c$-hadron contributions are separated by employing an additional tagging 
   algorithm~\cite{jpb}, which uses track impact parameters to select jets 
   with a small probability of originating from the primary vertex of the 
   event ({\sc jpb} tags). Sequential semileptonic decays are identified by 
   searching lepton-jets for the presence of additional soft leptons 
   ($e$ or $\mu$ with $p_T \geq 2 \; \gevc$) that are referred to as 
   {\sc slt} tags. In Ref.~\cite{apo}, we have used measured rates of 
   {\sc secvtx} and {\sc jpb} tags to determine the bottom and charmed 
   content of this data sample; we have then tuned the parton-level cross 
   sections predicted by the simulation, based upon the 
   {\sc herwig}~\cite{herwig} and {\sc qq}~\cite{cleo} Monte Carlo generator
   programs, to match the heavy-flavor content of the data. 
   Reference~\cite{apo} shows that rates of lepton- and away-jets with 
   {\sc secvtx} and {\sc jpb} tags, as well as the relevant kinematical
   properties of the data, can be modeled by tuning the simulation within 
   the theoretical and experimental uncertainties. However, the number of
   away-jets with {\sc slt} tags, which according to the simulation are 
   mostly due to $b\bar{b}$ production, is found to be significantly larger
   than what predicted by the conventional-QCD simulation. The observed
   discrepancy is consistent with previously reported 
   anomalies~\cite{derwent,yuntae,abbot}, and opens the possibility that 
   approximately 30\% of the presumed semileptonic decays of $b$- hadrons
   produced at the Tevatron is due to unconventional sources.

   Therefore, it is of interest to extend the earlier comparison to the
   yields of {\sc slt} tags contained inside lepton-jets. The present 
   analysis is based upon the same samples of data and simulated events 
   used in Ref.~\cite{apo}, and makes use of the same tuning of the 
   simulation. In Sec.~\ref{sec:ss-dilep}, we evaluate rates of lepton-jets
   containing also one soft lepton tag. In Sec.~\ref{sec:ss-kinslt}, we 
   compare the kinematics of these lepton pairs in the data and in the 
   simulation. Section~\ref{sec:ss-syst} contains cross-checks and a
   discussion of systematic effects. Our conclusions are presented in
   Sec.~\ref{sec:concl}.
%%%%%%%%%%%%%%%%%%%%%%%%%%%%%%%%%%%%%%%%%%%%%%%%%%%%%%%%%%%%%%%%
 \section{Lepton-jets containing an additional soft lepton}
 \label{sec:ss-dilep}
%%%%%%%%%%%%%%%%%%%%%%%%%%%%%%%%%%%%%%%%%%%%%%%%%%%%%%%%%%%%%%%%
   We search lepton-jets for additional soft leptons ($p_T \geq 2\; \gevc$)
   using the {\sc slt} algorithm~\cite{cdf-evidence,kestenbaum,cdf-tsig,suj}. 
   Pairs of trigger and soft leptons arise from four different sources:
   sequential semileptonic decays of single $b$ hadrons, leptonic decays of
   $\psi$ mesons, semileptonic decays of two different hadrons with heavy
   flavor produced by gluons branching into pairs of $b$ or $c$ quarks, and
   hadrons that mimic the experimental signature of a lepton. We compare data
   and simulation for the following yields of tags:
 \begin{enumerate}
 \item  $Dil$, the number of lepton-jets containing one and only one additional
        soft lepton. Since approximately 50\% of the $J/\psi$ mesons produced
        at the Tevatron do not originate from $B$ decays~\cite{psi-sec} and
        are not  modeled by the heavy flavor simulation, dileptons with
        opposite charge, same flavor, and invariant mass 
        $2.6 \leq m_{ee} \leq 3.6~\gevcc$ and 
        $2.9 \leq m_{\mu \mu} \leq 3.3~\gevcc$ are removed from this study.
 \item  $Dil^{SEC}$ ($Dil^{JPB}$), the number of lepton-jets that also 
        contain one soft lepton and a {\sc secvtx} ({\sc jpb}) tag due to 
        heavy flavor. Lepton pairs consistent with $J/\psi$ decays are 
        also removed.
 \end{enumerate}
   The yields of lepton pairs consistent with $J/\psi$ decays, which are
   removed from the present analysis, have been compared to the simulation
   in Table XIV of Ref.~\cite{apo}. The comparison has been used to verify
   the $b$ purity of the data.

   The observed numbers of jets containing a lepton pair are listed in 
   Table~\ref{tab:tab_rateslt_3}. Rates of dileptons produced by heavy 
   flavor decays in the simulation (as yet unnormalized) are shown in 
   Table~\ref{tab:tab_rateslt_4}. In the simulation, dileptons are mostly
   produced by sequential decays of single $b$ hadrons and have opposite 
   sign charge~(OS); approximately 5\% of the lepton pairs have same sign
   charge (SS) and are found in jets produced by gluons branching into pairs
   of heavy quarks.

   In the data the ratio of SS to OS dileptons is appreciably higher 
   ($\simeq$~20\%) than in the simulation. The excess of SS dileptons with
   respect to the simulation is attributed to hadrons that mimic the lepton
   signature. Therefore we use the number of SS dileptons with a 10\% error
   to estimate and remove the fake-lepton contribution to OS dileptons. 
   This rather intuitive method for estimating this background will be 
   further discussed in Section~\ref{sec:ss-syst}.
%%%%%%%%%%%%% Tables %%%%%%%%%%%%%%
  \begin{table}
 \caption{Numbers of lepton-jets containing an additional soft lepton. 
          OS and SS indicate lepton pairs with opposite and same sign
          charge, respectively. We use the difference OS-SS to remove 
          the contribution of fake leptons. Mistags are the numbers of
          dileptons in jets with fake {\sc secvtx} or {\sc jpb} tags. 
          The yields of lepton pairs ($Dil$) are also shown for different
          lepton flavors ($ee$, $\mu\mu$, and $e \mu$). }
 \begin{center}
 \begin{ruledtabular}
 \begin{tabular}{lcccclcccc}
  & \multicolumn{3}{c}{\bf Electron data} &  & 
  & \multicolumn{3}{c}{\bf Muon data} & \\
  Flavor                         &   OS    &  SS   & OS-SS  & Mistags  & 
  Flavor                         &   OS    &  SS   & OS-SS  & Mistags  \\
  $e e$                          & $~441$  & $107$ & $~334$ &          &
  $\mu \mu$                      &  $335$  & $107$ &  $228$ &          \\
  $e \mu$                        & $1009$  & $232$ & $~777$ &          &
  $\mu e$                        &  $141$  & $~33$ &  $108$ &          \\
  $Dil$ ($e e$+$e\mu$)           & $1450$  & $339$ & $1111$ &          &
  $Dil$ ($\mu \mu$+$\mu e$)      &  $476$  & $140$ &  $336$ &          \\
  $e e$                          & $~111$  & $~18$ & $~~93$ &          &
  $\mu \mu$                      &  $127$  & $~25$ &  $102$ &          \\
  $e \mu$                        & $~371$  & $~61$ & $~310$ &          &
  $\mu e$                        &  $~71$  & $~14$ &  $~57$ &          \\
  $Dil^{SEC}$ ($e e$+$e \mu$)    & $~482$  & $~79$ & $~403$ &  $2.0$   &
  $Dil^{SEC}$($\mu \mu$+$\mu e$) &  $198$  & $~39$ &  $159$ &  $0.8$   \\
  $e e$                          & $~143$  & $~20$ & $~123$ &          &
  $\mu \mu$                      &  $143$  & $~28$ &  $115$ &          \\
  $e \mu$                        & $~414$  & $~66$ & $~348$ &          &
  $\mu e$                        &  $~72$  & $~12$ &  $~60$ &          \\
  $Dil^{JPB}$ ($e e$+$e \mu$)    & $~557$  & $~86$ & $~471$ &  $7.0$   &
  $Dil^{JPB}$ ($\mu \mu$+$\mu e$)&  $215$  & $~40$ &  $175$ &  $3.5$   \\
 \end{tabular}
 \end{ruledtabular}
 \end{center}
 \label{tab:tab_rateslt_3}
 \end{table}

 %--------------------  Dileptons
 \begin{table}
 \caption{Number of lepton-jets containing an additional soft lepton due
          to heavy  ($b$ and $c$) quarks in the simulation not yet tuned
          according to the fit performed in Ref.~\protect\cite{apo}; 
          dir, f.exc, and GSP indicate the direct (LO), flavor excitation,
          and gluon splitting contributions predicted by {\sc herwig}
          to the numbers of OS/SS lepton pairs with different and same 
          flavor and to the numbers of OS-SS lepton pairs with any flavor.
          There is no contribution from $c$ direct production. At generator
          level, we have verified that both the trigger and soft lepton
          tracks match an electron or muon originating from $b$- or 
          $c$-hadron decays (including those coming from $\tau$ or $\psi$
          cascade decays). Lepton pairs consistent with $J/\psi$ decays
	  are removed as in the data.}
% \mediumtext
 \def\arraystretch{0.8}
 \begin{center}
 \begin{ruledtabular}
 \begin{tabular}{lccccc}
               & \multicolumn{5}{c}{\bf Electron simulation}  \\
 Flavor                & $b$ dir & $b$ f.exc & $c$ f.exc & $b$ GSP & $c$ GSP\\
 $ee$                  &  ~63/0  &  105/1    &   2/0     & ~57/12  &  12/0  \\
 $e\mu$                &  148/0  &  266/4    &   0/0     & 157/28  &  20/0  \\
 $Dil$ ($e e$+$e \mu$) &   211   &   367     &    2      &   174   &   32   \\
 $ee$                  &  ~24/0  &  ~39/1    &   0/0     & ~25/7~  &  ~0/0  \\
 $e \mu$               &  ~73/0  &  118/3    &   0/0     & ~61/8~  &  ~2/0  \\
 $Dil^{SEC}$ ($ee$+$e\mu$)&~97   &   153     &    0      &  ~71    &   ~2   \\
 $ee$                  &  ~31/0  &  ~61/1    &   0/0     & ~31/8~  &  ~4/0  \\
 $e \mu$               &  ~81/0  &  146/2    &   0/0     & ~85/13  &  ~2/0  \\
 $Dil^{JPB}$ ($ee$+$e\mu$)& 112  &   204     &    0      &  ~95    &   ~6   \\
              & \multicolumn{5}{c}{\bf Muon simulation} \\
  Flavor               & $b$ dir & $b$ f.exc & $c$ f.exc & $b$ GSP & $c$ GSP\\
 $\mu\mu$              & ~28/0   & ~65/2     &   3/1      & ~52/12 &  12/0  \\
 $\mu e$               & ~26/0   & ~27/0     &   0/0      & ~31/3  &  ~5/0  \\
 $Dil$ ($\mu\mu$+$\mu e$)& ~54   &  ~90      &    2       &  ~68   &   17   \\
 $\mu \mu$             & ~14/0   & ~35/2     &   1/0      & ~25/7~ &  ~0/0  \\
 $\mu e$               & ~21/0   & ~15/0     &   0/0      & ~16/1~ &  ~0/0  \\
 $Dil^{SEC}$ ($\mu\mu$+$\mu e$)& ~35& ~48    &    1       &  ~33   &    0   \\
 $\mu \mu$             & ~18/0   & ~38/2     &   1/0      & ~33/8~ &  ~5/0  \\
 $\mu e$               & ~19/0   & ~21/0     &   0/0      & ~18/2~ &  ~1/0  \\
 $Dil^{JPB}$ ($\mu \mu$+$\mu e$)&~37& ~57    &    1       &  ~41   &   ~6   \\
 \end{tabular}
 \end{ruledtabular}
 \end{center}
 \label{tab:tab_rateslt_4}
 \end{table}

 \newpage
%%%%%%%%%%%%%%%%%%%%%%%%%%%%%%%%%%%%%%%%%%%%%%%%%%%%%%%%%%%%%%%%%%%%%%%
 \subsection{Rates  of soft leptons due to heavy flavor in the data 
             and in the normalized simulation} \label{sec:ss-sltcomp}
%%%%%%%%%%%%%%%%%%%%%%%%%%%%%%%%%%%%%%%%%%%%%%%%%%%%%%%%%%%%%%%%%%%%%%%
   Table~\ref{tab:tab_rateslt_5} lists numbers of OS-SS lepton pairs 
   due to heavy flavor in the data and in the 
   simulation normalized according to the fit described in Sec.~VIII of 
   Ref.~\cite{apo}. Table~\ref{tab:tab_rateslt_6} lists the contribution
   of the various production mechanisms to the numbers of OS-SS lepton 
   pairs in the normalized simulation.

   In the data there are $1447 \pm 65$ lepton pairs in the same jet 
   (the statistical error is $\pm 43.8$ and the systematic uncertainty of
   the fake-lepton removal is $\pm 48.0$). The simulation predicts 
   $1180.7 \pm 128.7$ dileptons (the systematic uncertainty due to the 
   {\sc slt} tagging efficiency is $\pm 118$ and the uncertainty due to the
   fit used to tune the simulation in Ref.~\cite{apo} and to the simulation
   statistical error is $\pm 51.5$).

   In lepton-jets with {\sc jpb} tags, we find $635.5 \pm 30.5$ lepton pairs 
   (the statistical error is $\pm 27.8$ and the systematic uncertainty due 
   to the fake-lepton removal is $\pm 12.6$). The simulation predicts 
   $530.9 \pm 39.7$ lepton pairs (the systematic uncertainty due to the
   {\sc slt} tagging efficiency is $\pm 26.5$ and the uncertainty due to
   the tuning of the simulation and to the simulation statistical error 
   is $\pm 29.5$). The small excess of the data with respect to the 
   simulation is approximately a 2 $\sigma$ effect. In the next section,
   we study some kinematical properties of these lepton pairs.
%%%%%%%%%%%% Tables %%%%%%%%%%%%
 \newpage
  \begin{table}
 \caption{Number of OS-SS lepton pairs due to heavy flavors
          in the data and in the normalized simulation.}
 \begin{center}
 \def\arraystretch{1.2}
 \begin{ruledtabular}
 \begin{tabular}{lcccc}
   & \multicolumn{2}{c}{\bf Electrons}   & \multicolumn{2}{c}{\bf Muons} \\
  Tag type           &   Data            &    Simulation   
                     &   Data            &    Simulation      \\
 $Dil$               & $1111.0 \pm 54.2$ &  $893.0 \pm 101.3$ 
                     &  $336.0 \pm 28.5$ &  $287.7 \pm  38.6$ \\
 $Dil^{SEC}$         & $~401.0 \pm 25.0$ &  $366.6 \pm 32.1~$  
                     &  $158.2 \pm 15.9$ &  $143.0 \pm  17.6$ \\
 $Dil^{JPB}$         & $~464.0 \pm 26.8$ &  $387.3 \pm 32.0~$ 
                     &  $171.5 \pm 16.5$ &  $143.6 \pm  16.7$ \\
 \end{tabular}
 \end{ruledtabular}
 \end{center}
 \label{tab:tab_rateslt_5}
 \end{table}
%---------------

  \begin{table}
 \caption{Predicted numbers of OS-SS lepton pairs, listed by
          production mechanisms.}
 \begin{center}
 \def\arraystretch{1.0}
 \begin{ruledtabular}
 \begin{tabular}{lccccc}
            &  \multicolumn{5}{c}{\bf Electron simulation}               \\
    Tag type         &    $b$ dir      &    $b$ f.exc    &  $c$ f.exc 
                     &    $b$ GSP     &     $c$ GSP                     \\ 
 $Dil$               & $215.1\pm 28.6$ & $381.9\pm 56.4$ & $ 2.2\pm 1.7$ 
                     & $248.1\pm 49.0$ & $ 45.7\pm 14.5$                 \\
 $Dil^{SEC}$         & $100.1\pm 12.4$ & $161.7\pm 22.9$ & $ 0 $ 
                     & $102.5\pm 21.1$ & $ ~2.2\pm 1.9~$                 \\
 $Dil^{JPB}$         & $~93.9\pm 11.0$ & $175.0\pm 23.6$ & $ 0 $ 
                     & $111.4\pm 21.6$ & $ ~7.0\pm 3.4~$                 \\
            & \multicolumn{5}{c}{\bf Muon simulation}                    \\
    Tag type         &    $b$ dir      &    $b$ f.exc    &  $c$ f.exc 
                     &    $b$ GSP      &     $c$ GSP                     \\ 
 $Dil$               & $~58.1\pm 10.3$ & $~99.2\pm 17.8$ & $ 2.4\pm 2.4$ 
                     & $102.4\pm 23.0$ & $ 25.6\pm 9.3~$                 \\
 $Dil^{SEC}$         & $~38.2\pm 7.0~$ & $~53.6\pm 10.4$ & $ 0.9\pm 1.0$ 
                     & $~50.3\pm 12.9$ & $      0      $                 \\
 $Dil^{JPB}$         & $~32.8\pm 5.8~$ & $~51.6\pm 9.4~$ & $ 1.0\pm 1.0$ 
                     & $~50.8\pm 12.2$ & $ ~7.4\pm 3.6~$                 \\
 \end{tabular}
 \end{ruledtabular}
 \end{center}
 \label{tab:tab_rateslt_6}
 \end{table}

 \clearpage
%%%%%%%%%%%%%%%%%%%%%%%%%%%%%%%%%%%%%%%%%%%%%%%%%%%%%%%
 \section{Dilepton kinematics} \label{sec:ss-kinslt}
\begin{figure}
 \begin{center}
 \leavevmode
% \epsfxsize \textwidth
% \epsffile{kinslt_1.eps}
 \includegraphics*[width=\textwidth]{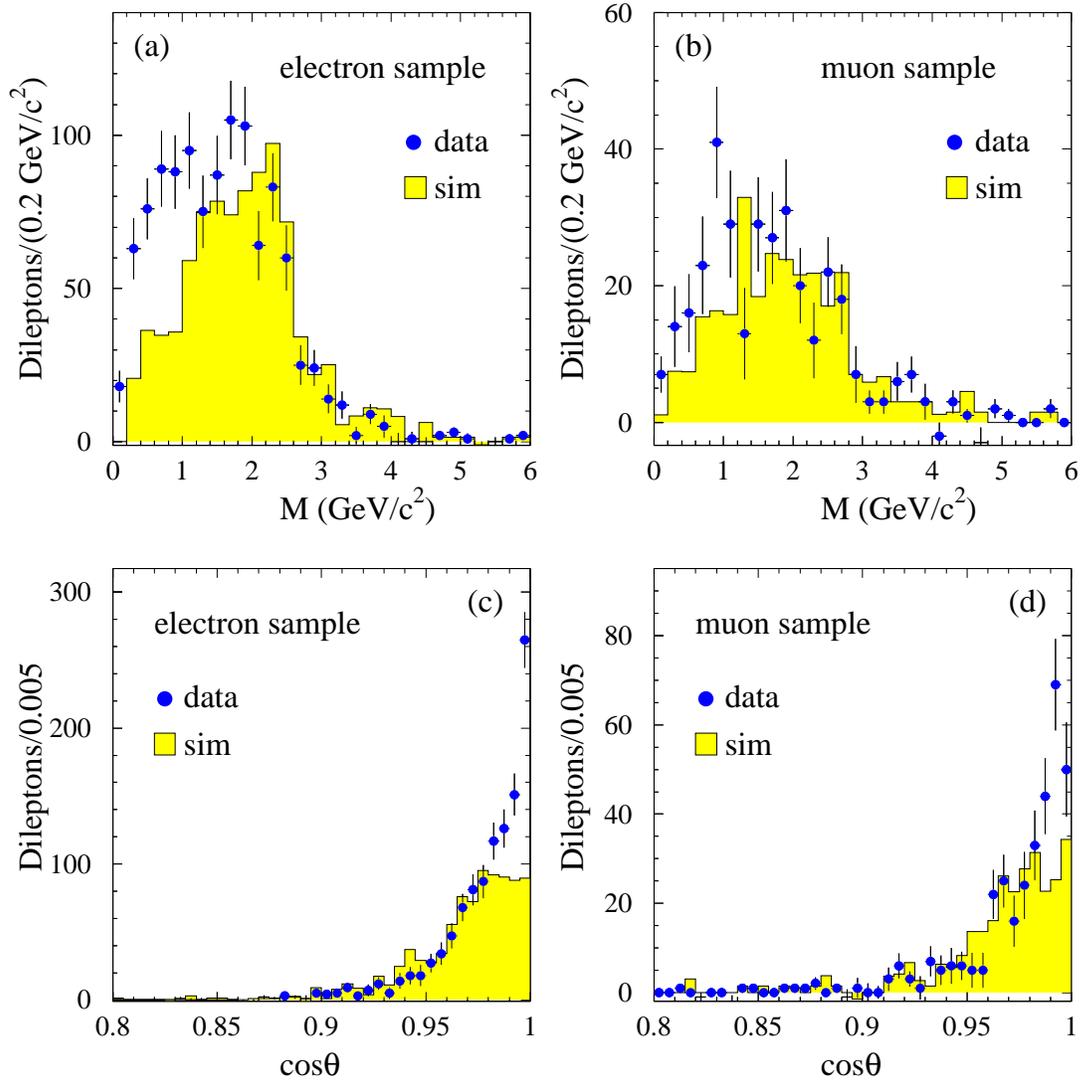}
 \caption[]{Dilepton invariant mass distributions in the inclusive 
            electron (a) and muon (b) samples are compared to the
            simulation prediction. Distributions of the dilepton 
            opening angle, $\theta$, in the two samples are shown 
            in (c) and (d).}
 \label{fig:kinslt_1}
 \end{center}
 \end{figure}

\begin{figure}
 \begin{center}
 \leavevmode
% \epsffile{kinslt_2.eps}
 \includegraphics{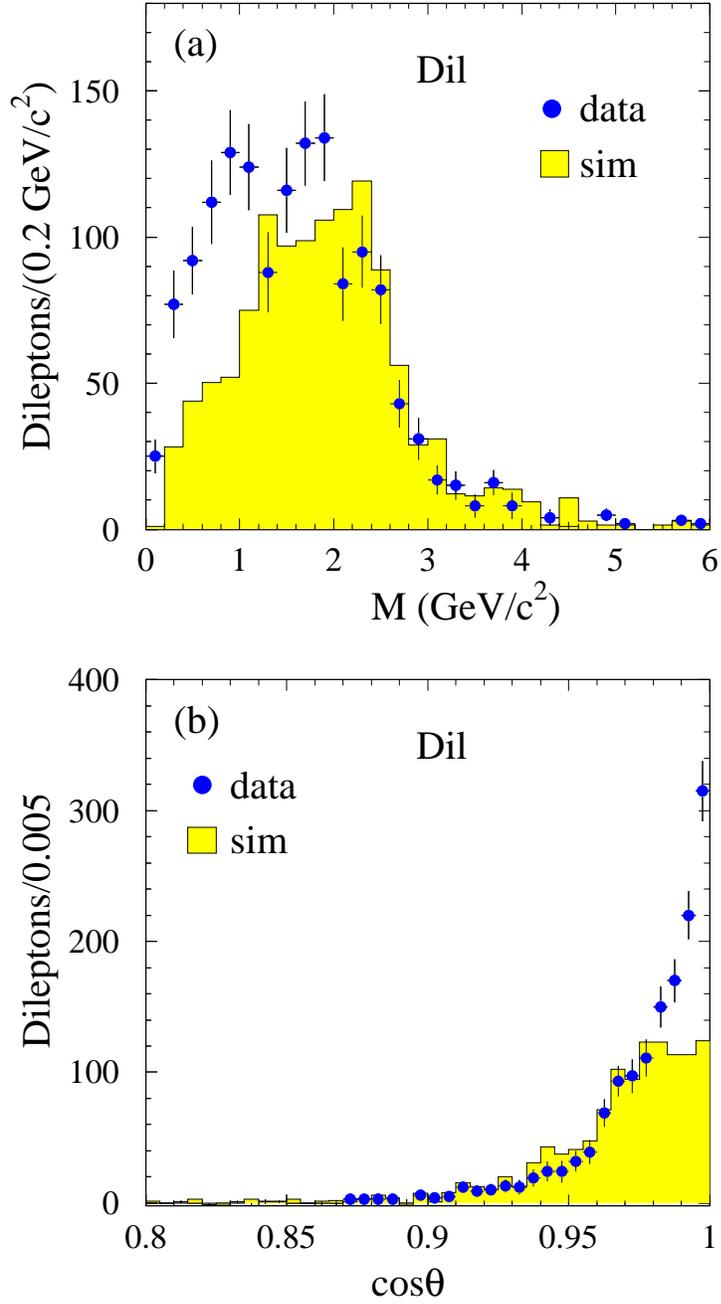}
 \caption[]{Distributions of the dilepton (a) invariant mass, $M$, and 
           (b) opening angle, $\theta$, for the inclusive lepton
           (electron+muon) sample and for its simulation.}
 \label{fig:kinslt_2}
 \end{center}
 \end{figure}

\begin{figure}
 \begin{center}
 \leavevmode
% \epsfxsize \textwidth
% \epsffile{kinslt_3.eps}
 \includegraphics*[width=\textwidth]{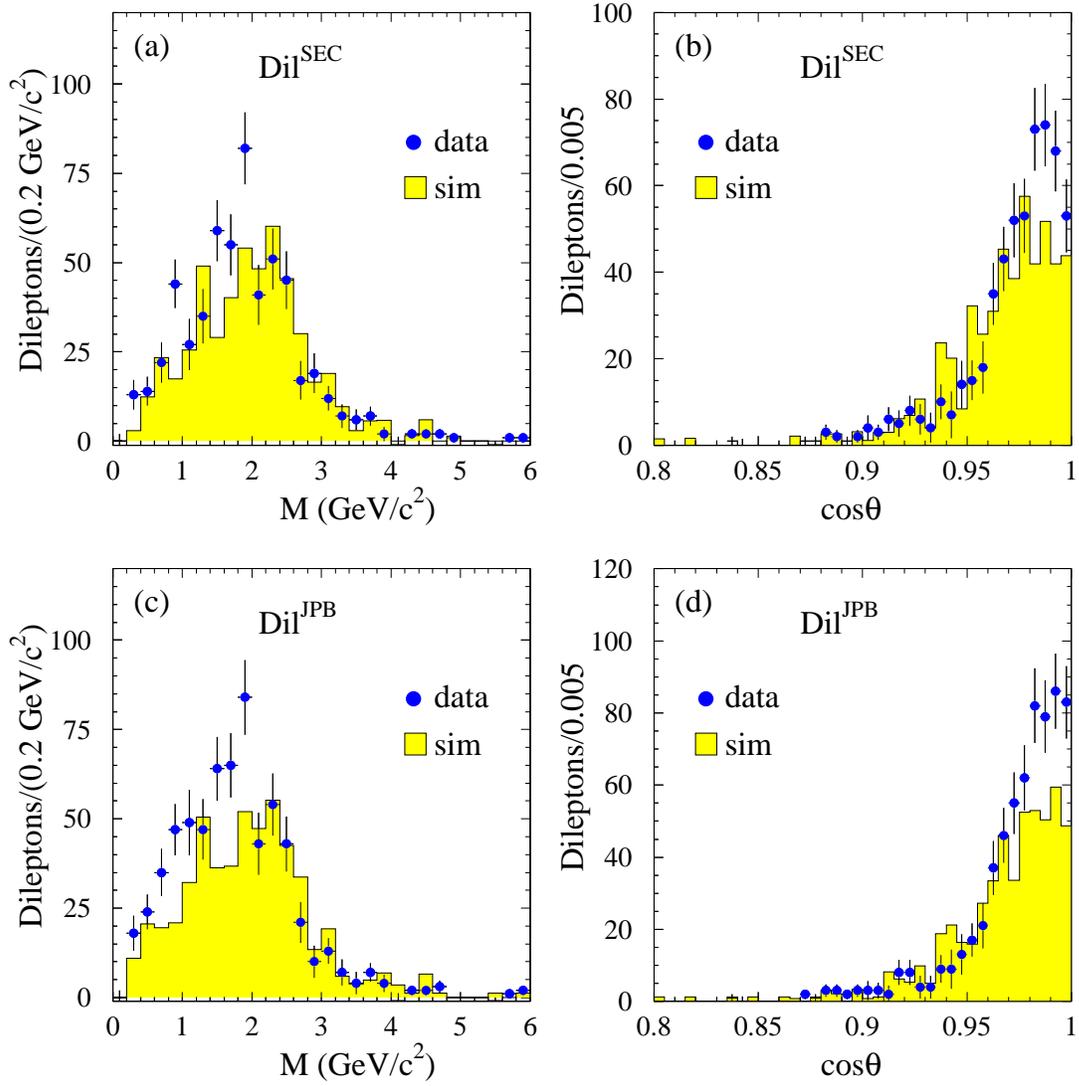}
 \caption[]{Distributions of (a) invariant mass and (b) opening angle
            of lepton pairs contained in jets tagged by the {\sc secvtx} 
            or (c and d) {\sc jpb} algorithms for the inclusive 
            lepton sample and its simulation.}
 \label{fig:kinslt_3}
 \end{center}
 \end{figure}

\begin{figure}
 \begin{center}
 \leavevmode
% \epsffile{kinslt_4.eps}
 \includegraphics{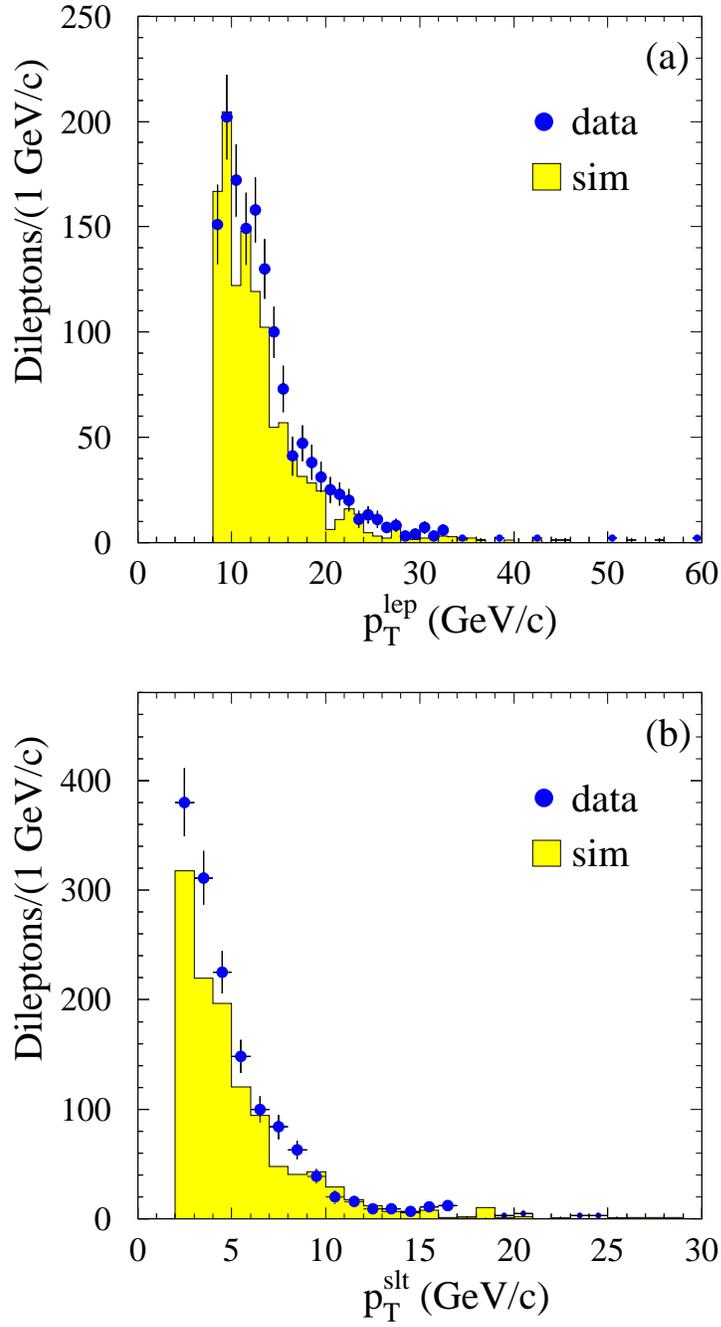}
 \caption[]{Transverse momentum distributions of the trigger lepton 
            (lep) and the accompanying soft lepton (slt).}
 \label{fig:kinslt_4}
 \end{center}
 \end{figure}

%%%%%%%%%%%%%%%%%%%%%%%%%%%%%%%%%%%%%%%%%%%%%%%%%%%%%%%
   Figures~\ref{fig:kinslt_1} to~\ref{fig:kinslt_3} compare distributions
   of invariant mass and opening angle of dileptons contained in the same
   jet in the data and in the simulation~\footnote{ 
   The fake-lepton background is removed by subtracting the distribution 
   of SS dileptons from that of OS dilepton, both in the data and in the 
   simulation. In the data, errors include the $\pm 10$\% systematic 
   uncertainty of this removal.}. 
   The small excess of lepton pairs with respect to the simulation prediction
   (see Table~\ref{tab:tab_rateslt_5}) appears to be concentrated at 
   invariant masses smaller than 2 $\gevcc$ and opening angles smaller 
   than 0.2 rad. For dilepton invariant masses larger than 2 $\gevcc$ data
   and simulation are in reasonable agreement. The shapes of the transverse
   momentum distributions of the trigger and soft leptons, shown in 
   Fig.~\ref{fig:kinslt_4}, are compatible with that of the expectation.
\newpage
%%%%%%%%%%%%%%%%%%%%%%%%%%
%%%%%%%%%%%%%%%%%%%%%%%%%
%%%%%%%%%%%%%%%%%%%%%%%%%%
%%%%%%%%%%%%%%%%%%%%%%%%%
%%%%%%%%%%%%%%%%%%%%%%%%%
%%%%%%%%%%%%%%%%%%%%%%%%%
%%%%%%%%%%%%%%%%%%%%%%%%%
%%%%%%%%%%%%%%%%%%%%%%%%%
%%%%%%%%%%%%%%%%%%%%%%%%%%%%%%%%%%%%%%%%%%%%%%
 \clearpage
 \section{Systematics}\label{sec:ss-syst}
%%%%%%%%%%%%%%%%%%%%%%%%%%%%%%%%%%%%%%%%%%%%%%
   The excess of lepton pairs with respect to the simulation is all 
   concentrated at dilepton opening angles smaller than $11 \deg$. Since 
   this is approximately the angle covered by a central calorimeter tower,
   we have investigated at length the possibility that the efficiency of the
   lepton selection criteria, described in Sec. IV of Ref.~\cite{apo}, is
   not simulated properly when two lepton-candidate
   tracks hit the same calorimeter tower. However, for opening angles 
   smaller than $11 \deg$, the excess with respect to the simulation when
   the leptons are contained in the same tower is smaller, but consistent
   with that observed when the leptons hit two neighboring towers. We have
   also inspected all distributions of the tracking and calorimeter 
   informations that are used to select leptons. We have compared these 
   distributions for lepton pairs with opening angle smaller and larger 
   than $11 \deg$ without discovering any sensible difference. Therefore,
   we have investigated other possible causes; since they are of general
   interest, we present these studies in the following. In subsection~A 
   we verify the method used to estimate and remove the fake-lepton 
   contribution. In subsection~B we check the simulation of sequential 
   $b$-decays that represents the largest contribution to lepton pairs. 
   Finally, in subsection~C we study a handful of events containing
   three leptons. 
%%%%%%%%%%%%%%%%%%%%%%%%%%%%%%%%%%%%%
 \subsection{Fake-lepton estimate}
%%%%%%%%%%%%%%%%%%%%%%%%%%%%%%%%%%%%%
   The technique of removing  the fake-lepton background by subtracting SS
   dileptons from OS dileptons has been used by CDF in several measurements
   of the Drell-Yan cross section~\cite{drellyan}. We prefer this technique
   to the use of the standard parametrized probability of finding a fake
   lepton in a jet, derived using large samples of generic-jet 
   data~\cite{apo,cdf-tsig,suj}, because the latter method might not 
   be applicable to jets that already contain a lepton (generic-jet data
   do not contain enough lepton pairs to construct a reliable parametrization
   of this fake probability).

   The simulated inclusive electron sample contains $955\pm 108$ OS and 
   $63\pm 9$ SS dileptons produced by heavy-quark decays. In the data, there 
   are $1450$ OS and $339$ SS dileptons. After removing the dilepton rates 
   predicted by the simulation, we would like to explain in terms of
   fake-lepton background the remaining $495\pm 114$ OS and $276\pm 20$ SS
   dileptons~\footnote{
   The standard simulation ignores $b$-hadron mixing and underestimates
   the rate of SS dileptons due to the decays of two different $b$
   hadrons. This is a small effect; when using the time-integrated mixing
   parameter $\bar{\chi}=0.118$~\cite{pdg}, the simulation predicts 
   $915\pm 105$ OS and $103\pm 15$ SS dileptons produced by heavy-quark 
   decays. Therefore, the difference between data and simulation, which 
   should be attributed to the fake-lepton background, becomes
   $535\pm 111$ OS and $236\pm 24$ SS dileptons.}.
 
   In principle, rates of OS and SS dileptons due to misidentification
   background could be different. On average, jets contain the same number 
   of positive and negative lepton-candidate tracks. Therefore, when 
   searching for an additional soft lepton jets in which one track has been
   already identified as the trigger lepton, the number of OS candidates is
   larger than the number of SS candidates (these numbers will be 
   approximately equal only for jets with a very large number of candidate
   tracks).

   We have investigated this scenario by using samples of generic jets 
   (JET~20, JET~50, and JET~70) and their simulation described in 
   Refs.~\cite{apo,cdf-tsig,suj}; the simulation has been also tuned to 
   reproduce the rates of {\sc secvtx} and {\sc jpb} tags observed in the 
   data. We select jets containing an {\sc slt} tag, and for these jets 
   we count the number of additional {\sc slt} candidate tracks, $N_{C}$, 
   with opposite or same charge. We also count the number of OS and SS
   additional soft lepton tags, $Dil$, found in these jets. These rates 
   are listed in Table~\ref{tab:tab_syst_1}. As expected, the table shows
   a large difference between the number $N_C$ of OS and SS candidates.
   However, in generic jets, which are not rich in heavy flavor, the rate
   of OS and SS soft lepton pairs inside the same jet is approximately equal 
   (to within 13\%). After removing the heavy flavor contribution predicted
   by the simulation, we derive $P_{fk}$, the probability of converting
   {\sc slt} candidate tracks into a fake {\sc slt} tag in jets that already
   contain an {\sc slt} tag (see Table~\ref{tab:tab_syst_1}). We note that
   in generic-jet data the probability $P_{fk}$ for OS candidates is 65\% 
   of that for SS candidates.

   Our standard estimate of the number of OS dileptons due to
   misidentification background assumes that it is equal to the number
   of observed SS dileptons minus the number of predicted SS dileptons
   due to heavy flavor [$339-(63\pm 9)= 276\pm 20$]. We use two additional
   methods to verify this estimate.

   In the inclusive electron sample, e-jets contain $N_c(OS)=54938$ and 
   $N_c(SS)=34744$ candidate tracks, respectively. In the first method, 
   we multiply the numbers of candidates by the corresponding probabilities 
   $P_{fk}$ derived in generic-jet data. We predict a slightly smaller OS
   background ($236\pm 38$) and an amount of SS background ($229\pm 14$)
   in agreement with the estimate that includes $b$-hadron mixing~$^2$.

   An additional estimate of the OS background can be derived by applying
   the standard parametrization~\cite{apo,cdf-tsig,suj} of the fake {\sc slt} 
   probability to all OS candidate tracks. This method yields a slightly 
   higher background estimate of $302\pm 30$ OS fake dileptons~\footnote{
   This method for estimating the fake-lepton contribution has been
   previously used by CDF in a measurement~\cite{wendy} of the $b$-quark 
   fragmentation fractions of strange and light $B$ mesons by using dimuons
   with invariant mass $m_{\mu\mu} \leq 2.8\; \gevcc$.}.

   We take the $\pm 10$\% discrepancy between the different background 
   estimates as a measure of its uncertainty. We note that the simulation
   of the {\sc slt} algorithm relies on parametrizations based on the
   data and does not provide a good understanding of why, in generic-jet
   data, the probability $P_{fk}$ is smaller for OS candidates than for
   SS candidates. However, imposing the conditions that the excess of
   $535\pm 111$ OS
   pairs with respect to the simulation prediction~$^2$ is 
   due to fake-lepton background and that the probability $P_{fk}$ does 
   not depend on the pair sign produces a paradoxical result. In the
   inclusive electron sample, all $339$ SS pairs are attributed to fake
   background whereas the simulation predicts $103 \pm 15$ SS pairs due
   to heavy flavor; in contrast, generic jets require the presence of
   $73 \pm 25$ SS pairs due to heavy flavor, not predicted by the 
   simulation and of the same size of the predicted number of OS pairs 
   due to heavy flavor ($63 \pm 24$ in Table~\ref{tab:tab_syst_1}).

\begin{figure}
 \begin{center}
 \vspace{-0.4in}
 \leavevmode
% \epsffile{fig_syst_5.eps}
 \includegraphics{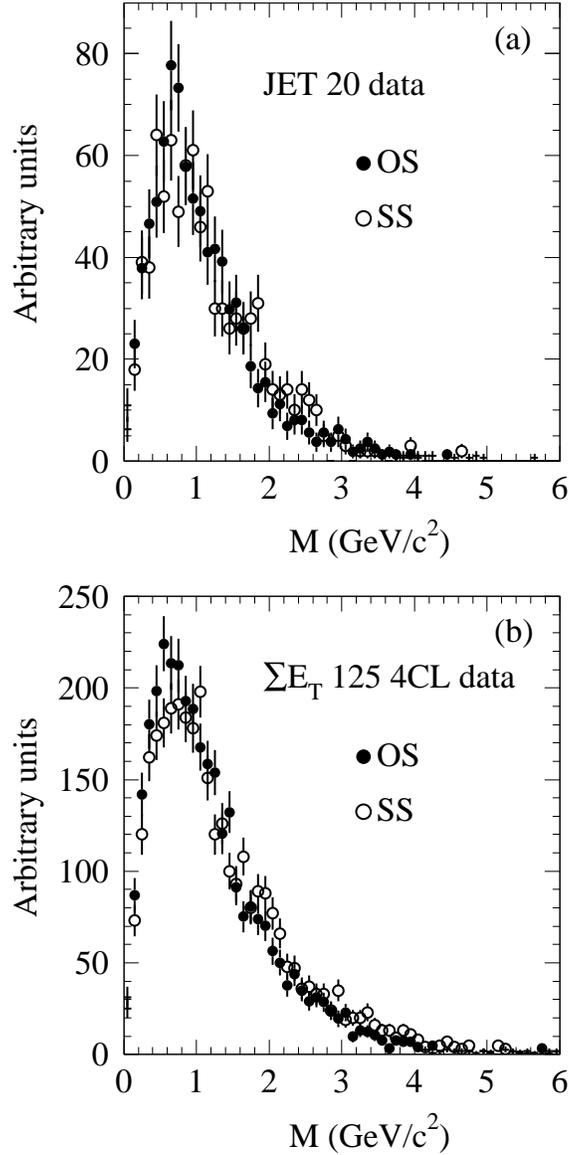}
 \caption[]{Invariant mass distributions between a track with 
            $p_T\geq 8\; \gevc$ and a soft lepton track contained in the
            same jet; OS and SS distributions are normalized to the same area.}
 \label{fig:fig_syst_5}
 \end{center}
 \end{figure}

   We use the generic-jet data to also verify that OS and SS dileptons due
   to misidentification background have similar invariant mass distributions.
   For this purpose, we use two data sets, the jets of which have an average
   transverse energy comparable to that of jets in the inclusive lepton
   samples. The first data set is selected requiring the presence of at 
   least one jet with transverse energy larger than $20$ GeV (JET~20). 
   The second data set is selected requiring the presence of at least four
   jets with transverse energy larger than $15$ GeV and total transverse
   energy larger than $125$ GeV ($\sum E_T$~125~4CL). In order to emulate 
   the inclusive lepton sample requirement of one lepton with transverse
   momentum larger than $8\; \gevc$, we select jets with at least one track
   with $p_T \geq 8\; \gevc$ inside a cone of radius 0.4 around their axis. 
   We then search these jets for soft lepton tags. Figure~\ref{fig:fig_syst_5}
   shows that the invariant mass distributions of the high $p_T$ track and
   the soft lepton track for OS and SS combinations are indeed quite similar.
%%%%%%%%%%%% Table %%%%%%
  \begin{table}
 \caption{Number of tracks candidate to become soft lepton tags, $N_C$,
          in jets which already contain a soft lepton tag. $Dil$ is the
          number of jets containing two soft lepton tags. $Dil(h.f.)$ is the 
          number of jets containing dileptons produced by heavy quark
          decays predicted by the simulation. The difference between data
          and simulation prediction, $Dil(fk)$, is attributed to fake 
          dileptons. $P_{fk}=(Dil-Dil(h.f.))/N_C$ is the probability that 
          a track produces a fake soft lepton tag.}
 \begin{center}
 \begin{ruledtabular}
 \begin{tabular}{cccccc}
     & $N_{C}$   & $Dil$ & $Dil(h.f.)$ & $Dil(fk)$ &$P_{fk}$ \\
     \multicolumn{6}{c}{ JET~20} \\
  OS  & ~7684 & ~44 & 17.7 $\pm$ 7.9~ &~26.3 $\pm$ 10.3 & 0.0034 $\pm$ 0.0013\\
  SS  & ~5606 & ~40 &        0        &       ~40       & 0.0071 $\pm$ 0.0011\\
     \multicolumn{6}{c}{ JET~50} \\
  OS  & 14467 & ~85 & 21.0 $\pm$ 10.3 &~64.0 $\pm$ 13.8 & 0.0044 $\pm$ 0.0009\\
  SS  & 11460 & ~71 &        0        &       ~71       & 0.0062 $\pm$ 0.0007\\
    \multicolumn{6}{c}{ JET~70} \\
  OS  & 18329 & 107 & 23.8 $\pm$ 11.7 &~83.2 $\pm$ 15.6 & 0.0045 $\pm$ 0.0008\\
  SS  & 14666 & ~97 &        0        &       ~97       & 0.0066 $\pm$ 0.0006\\
   \multicolumn{6}{c}{ Sum} \\
  OS  & 40480 & 236 & 62.6 $\pm$ 23.8 &173.4 $\pm$ 28.3 & 0.0043 $\pm$ 0.0007\\
  SS  & 31732 & 208 &        0        &       208       & 0.0066 $\pm$ 0.0004\\
 \end{tabular}
 \end{ruledtabular}
 \end{center}
 \label{tab:tab_syst_1}
 \end{table}

%%%%%%%%%%%%%%%%%%%%%%%%%
 \clearpage
%%%%%%%%%%%%%%%%%%%%%%%%%%%%%%%%%%%%%%%%%%%%%%%%%%%%%%%%%%%%%%%%%%%%%%%%%%%
 \subsection{Simulation of sequential $b$-decays}\label{sec:ss-delphi}
%%%%%%%%%%%%%%%%%%%%%%%%%%%%%%%%%%%%%%%%%%%%%%%%%%%%%%%%%%%%%%%%%%%%%%%%%%%
   As shown in Table~\ref{tab:tab_rateslt_6}, most
   lepton pairs contained in the same jet  are produced by sequential decays
   of single $b$-hadrons. Sequential decays of $b$ hadrons are modeled with 
   the CLEO Monte Carlo generator ({\sc qq})~\cite{cleo}. The discrepancy
   in the shape of the invariant mass and opening angle distributions in the
   data and in the simulation could be due to an inaccurate modeling of these
   decays. We cannot verify this by using any of our data, and we do it 
   through a comparison of the process $e^{+}e^{-} \rightarrow b\bar{b}$ at 
   the $Z$-pole using the {\sc jetset}~7.3 Monte Carlo generator~\cite{jetset}
   as implemented at LEP by the DELPHI collaboration~\cite{lepar} and our
   simulation which uses the {\sc herwig} and {\sc qq} generators. At our 
   request, the DELPHI collaboration has compared dilepton invariant mass 
   distributions in hadronic $Z$-decays to their simulation using selection
   criteria that could be easily reproduced in the CDF detector~\cite{delphi}.

\begin{figure}
 \begin{center}
 \vspace{-0.4in}
 \leavevmode
% \epsfxsize \textwidth
% \epsffile{fig_syst_6.eps}
 \includegraphics*[width=\textwidth]{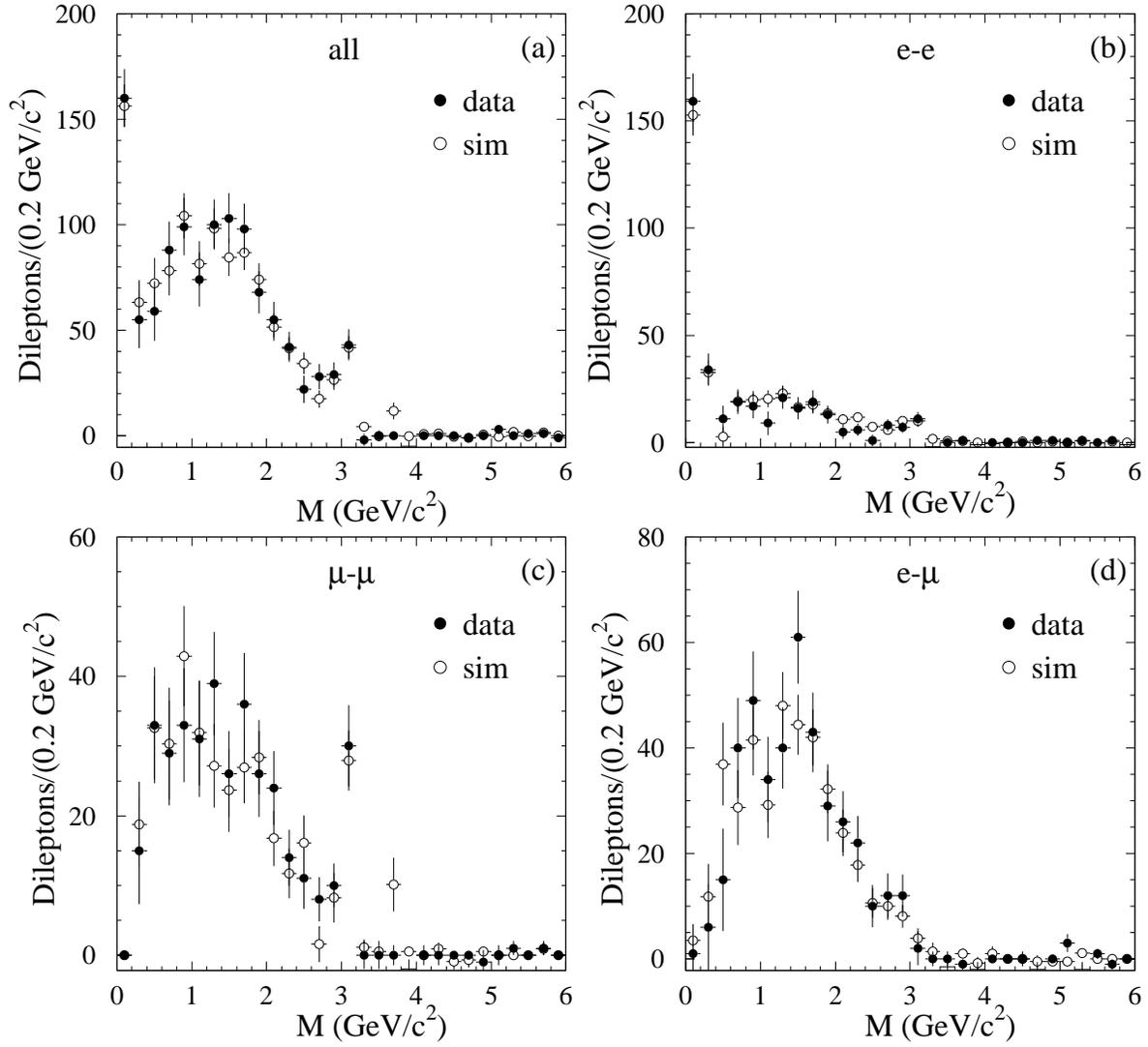}
 \caption[]{Invariant mass distributions of all OS-SS lepton pairs identified
            in events due to hadronic $Z$-decays ($\bullet$) by the DELPHI
            experiment at LEP are compared to the DELPHI simulation based on
            the {\sc jetset}~7.3 generator (open circle) for (a) all 
            dileptons, (b) $ee$ pairs, (c) $\mu \mu$ pairs, and 
            (d) $e \mu$ pairs.}
 \label{fig:fig_syst_6}
 \end{center}
 \end{figure}

\begin{figure}
 \begin{center}
 \vspace{-0.4in}
 \leavevmode
%\epsfxsize \textwidth
% \epsffile{fig_syst_7.eps}
 \includegraphics{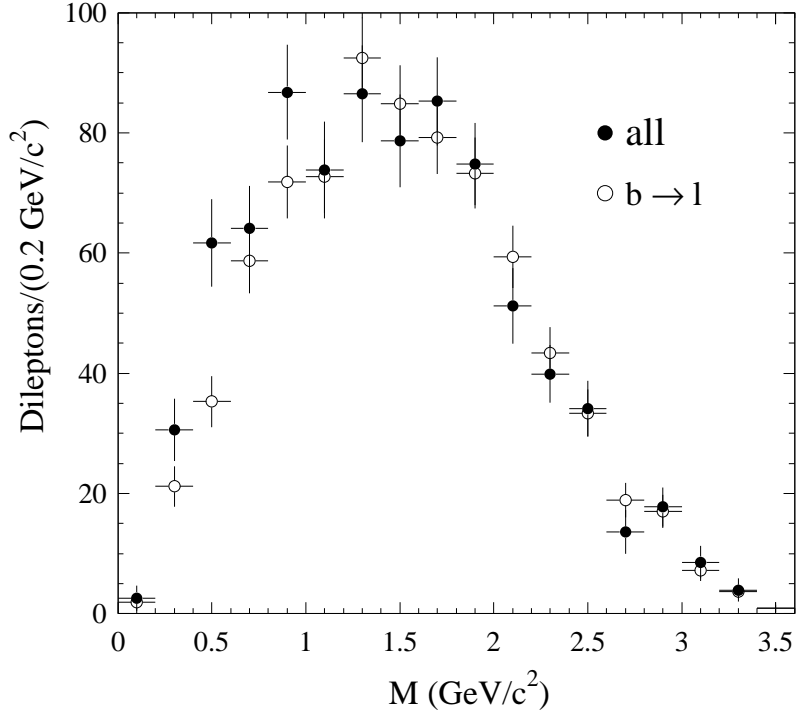}
 \caption[]{The invariant mass distribution of all OS-SS lepton pairs 
            identified in the DELPHI simulated sample of 
            $Z\rightarrow b\bar{b}$ events is compared to the distribution
            of OS-SS dileptons identified at generator level as coming from 
            $b$-quark decays. As in our study, $J/\psi$ mesons have been
            removed.}
 \label{fig:fig_syst_7}
 \end{center}
 \end{figure}

\begin{figure}
 \begin{center}
 \vspace{-0.4in}
 \leavevmode
 %\epsffile{fig_syst_8.eps}
 \includegraphics{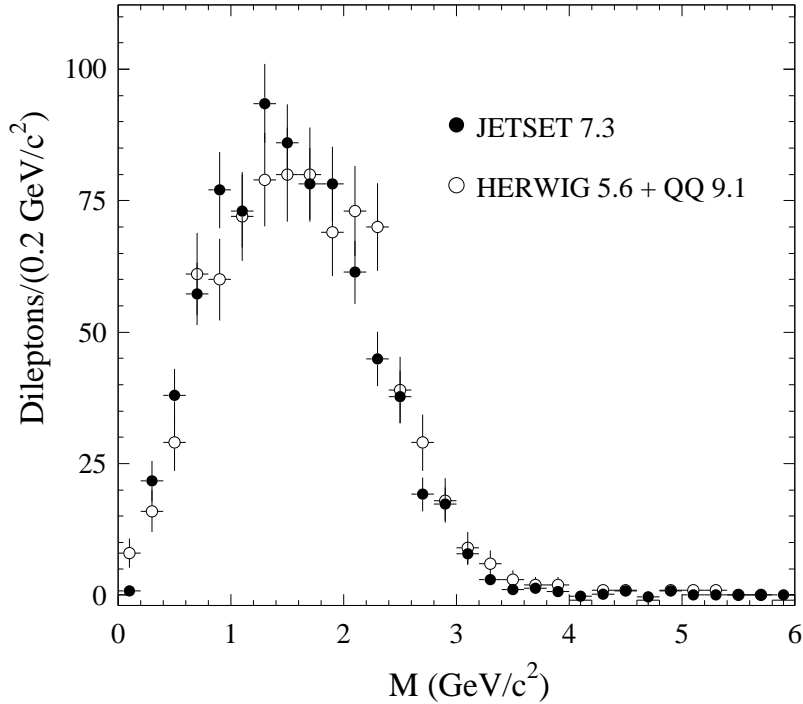}
 \caption[]{Invariant mass distributions of OS-SS dileptons in a simulation
            of the process $e^{+} e^{-} \rightarrow b\bar{b}$ at the $Z$-pole
            using the {\sc jetset}~7.3 generator and the generator package 
            {\sc herwig}~5.6~+~{\sc qq}~9.1.}
 \label{fig:fig_syst_8}
 \end{center}
 \end{figure}

   The DELPHI data consist of $573474$ hadronic $Z$-decays and the simulation
   of $992988$ events which are normalized to the same luminosity of the data.
   The DELPHI samples consist of events with $|cos(\theta_{T})| \leq 0.95$,
   where $T$ is the thrust axis~\cite{thrust}. The two leptons ($e$ or $\mu$)
   are required to belong to the same hemisphere, as defined by the thrust
   axis. Leptons are selected with pseudorapidity $|\eta| \leq 1$ and 
   momenta larger than $3\; \gevc$ to mimic closely the selection criteria of
   the {\sc slt} algorithm used by CDF. Dileptons are divided into OS and SS
   pairs. The comparison of the invariant mass distribution of OS-SS dileptons
   in the DELPHI data and simulation is shown in Fig.~\ref{fig:fig_syst_6}. 
   The {\sc jetset} generator models correctly the data at the $Z$-pole,
   where $b$ quarks are produced with an energy comparable to that in our
   inclusive lepton sample; the DELPHI simulation correctly models all 
   fake-lepton backgrounds including photon conversions, which are clearly
   visible in Fig.~\ref{fig:fig_syst_6}(a). CDF identifies and removes
   conversions in the  data and in the simulation, and then uses SS pairs 
   to estimate the remaining fake-lepton background. Our technique for
   removing misidentification background is supported by the plot in 
   Fig.~\ref{fig:fig_syst_7}.  For $Z \rightarrow b\bar{b}$
   events generated with the DELPHI simulation,
   the invariant mass distribution of all
   OS-SS lepton pairs is almost identical to that
   for OS-SS dilepton pairs identified
   at generator level as coming from $b$  decays ($J/\psi$ mesons are removed
   as in our analysis).
%%%%%%%%%%%%%%%%%%%%%%%%%%
%%%%%%%%%%%%%%%%%%%%%%%%%
%%%%%%%%%%%%%%%%%%%%%%%%%
 \clearpage
%%%%%%%%%%%%%%%%%%%%%%%%%
   Figure~\ref{fig:fig_syst_8} compares invariant mass distributions of OS-SS 
   dileptons in a simulation of the process $e^+ e^{-} \rightarrow b\bar{b}$
   at the $Z$-pole using the {\sc jetset}~7.3 generator and the DELPHI
   detector simulation and in a simulation of the same process that uses the
   generator package {\sc herwig}~5.6+{\sc qq}~9.1 and the CDF detector
   simulation. Dileptons are identified at generator level as coming from 
   $b$ decays ($J/\psi$ mesons are excluded). Given the fair agreement between
   the two event generators, it is unlikely that a simulation deficiency is
   responsible for the large discrepancy between the observed and predicted
   shapes of the invariant mass and opening angle distributions of lepton
   pairs arising from sequential decays of $b$ hadrons produced at the
   Tevatron (see Fig.~\ref{fig:kinslt_2}).
%%%%%%%%%%%%%%%%%%%%%%%%%%
%%%%%%%%%%%%%%%%%%%%%%%%%%%%%%%%%%%%%%%%%%%%%%%%%%%%%%%%%%%%%%%
 \subsection{Events with three leptons} \label{sec:ss-3slt}
%%%%%%%%%%%%%%%%%%%%%%%%%%%%%%%%%%%%%%%%%%%%%%%%%%%%%%%%%%%%%%%
   Events in which the lepton-jet contains an additional {\sc slt} tag 
   and the away-jet also contains a soft lepton tag are the most interesting
   because the rate of a-jets with {\sc slt} tags is also larger than the 
   prediction~\cite{apo}. Unfortunately, there is only a handful of these 
   events. As shown in Fig.~\ref{fig:kinslt_2}, the dilepton opening angle
   distribution is correctly modeled by the simulation for
   $\cos \theta \leq 0.975$; in contrast, only 60\% of the events is
   accounted for by the simulation for $\cos \theta \geq 0.975$. We compare
   the fraction of {\sc slt} tags in away-jets recoiling a jet containing
   a lepton pair with opening angle $\cos \theta \geq 0.975$ and
   $\cos \theta  \leq 0.975$ to verify if there is a correlation between 
   the excess of dilepton pairs and the excess of {\sc slt} tags in the
   away-jets. Rates of events with three leptons are listed in 
   Table~\ref{tab:tab_3slt_1}. The fraction of away-jets with {\sc slt}
   tags is ($2.34\pm 0.76$)\% for $\cos \theta \geq 0.975$ and
   ($1.07\pm 1.07$)\% for $\cos \theta  \leq 0.975$. The two fractions are
   statistically compatible, however a large correlation is also not excluded.
%%%%%%%%%% Table %%%%%%%%
  \begin{table}
 \caption{Numbers of away-jets recoiling against a lepton-jet that 
          also contains a soft lepton, $N_{ a-jet}$. $N_{ a-jet}^{ SLT}$ is
          the number of away-jets with an {\sc slt} tag due to heavy flavor;
          the listed number of fake {\sc slt} tags (mistags) has been removed.
          The numbers of a-jets are split according to $\theta$, the opening
          angle of the lepton pair ($Dil$). The SS dilepton contributions are
          used to remove the fake-lepton background to OS lepton pairs.}
 \begin{center}
 \begin{ruledtabular}
 \begin{tabular}{lcccc}
 \multicolumn{5}{c}{\bf Electron data}\\
      & \multicolumn{2}{c} {$\cos \theta \geq 0.975$ }
      & \multicolumn{2}{c} {$\cos \theta \leq 0.975$ } \\
  $Dil$          &  $N_{a-jet}$                & $N^{SLT}_{a-jet}$ (mistags) & 
  $N_{a-jet}$    &  $N^{SLT}_{a-jet}$ (mistags)  \\
   OS            &      1010                   &    22.8 (15.2)              &
      560        &     8.7 (8.3)     \\
   SS            &       198                   &    ~6.4 (2.6)~              &
      170        &     5.5 (2.5)     \\
  OS-SS          &    812 $\pm$ 34.7           &    16.4 $\pm$ 6.6           &
  390 $\pm$ 27.0 &    3.2 $\pm$ 5.0  \\
 \multicolumn{5}{c}{\bf Muon data}   \\
      & \multicolumn{2}{c} {$\cos \theta \geq 0.975$ }
      & \multicolumn{2}{c} {$\cos \theta \leq 0.975$ } \\
  $Dil$          &  $N_{a-jet}$                & $N^{SLT}_{a-jet}$ (mistags) &
  $N_{a-jet}$    &  $N^{SLT}_{a-jet}$ (mistags) \\
   OS            &      ~330                   &     10.6 (5.4)~             &
      191        &     2.7 (3.3)     \\
   SS            &      ~~94                   &      2.5 (1.5)~             &
      ~61        &     0.3 (0.7)     \\
  OS-SS          &    236 $\pm$ 20.6           &    ~8.1 $\pm$ 4.4           &
  130 $\pm$ 15.1 &    2.4 $\pm$ 2.6   \
 \end{tabular}
 \end{ruledtabular}
 \end{center}
 \label{tab:tab_3slt_1}
 \end{table}

 \clearpage
%%%%%%%%%%%%%%%%%%%%%%%%%%%%%%%%%%%%%%%%%%
 \section{Conclusions} \label{sec:concl}
%%%%%%%%%%%%%%%%%%%%%%%%%%%%%%%%%%%%%%%%%%
   We have studied rates and kinematical properties of sequential semileptonic
   decays of single $b$ hadrons produced at the Fermilab Tevatron collider.
   This study completes the review of the heavy flavor properties of jets
   produced at the Tevatron reported in Ref.~\cite{apo}. As in the previous
   analysis, we use events with two or more central jets with 
   $E_T\geq 15$ GeV, one of which (lepton-jet) is consistent with a
   semileptonic bottom or charmed decay to a lepton with $p_T\geq 8\; \gevc$.
   In the previous study, we have used measured rates of lepton- and away-jets
   containing displaced vertices ({\sc secvtx} tags) or tracks with large
   impact parameter ({\sc jpb} tags) to determine the bottom and charmed 
   content of the data; we have then tuned the parton-level cross section 
   predicted by the {\sc herwig}-based simulation
   to match the observed heavy-flavor content.
   The  simulation, tuned within the experimental and theoretical
   uncertainties, models correctly  rates of lepton- and away-jets with 
   {\sc secvtx} and {\sc jpb} tags, as well as the relevant kinematical
   properties of the data;  however, it underestimates by 50\% the number of
   away-jets containing a soft ($p_T \geq 2\; \gevc$) lepton.
 
   The present study uses the same tuning of the simulation, and extends the
   comparison to the rates of soft  leptons in the
   lepton-jet. We compare rates of jets containing a lepton pair (the trigger
   lepton and a soft lepton) with opposite (OS) and same (SS) sign charge
   in the data and in the conventional-QCD simulation.
   The data have a $20$\% excess of OS-SS
   dileptons with respect to the simulation ($2\sigma$ systematic effect).
   The distributions of the dilepton invariant mass and opening angle in
   the data are strikingly different from the simulation prediction for
   the observed excess is all concentrated at dilepton invariant masses
   smaller than $2\; \gevcc$ and opening angles smaller than $11^{\deg}$.
 
   Since all QCD-based simulations predict that rates of OS-SS dileptons
   with small opening angle are dominated by sequential semileptonic decays
   of single $b$-hadrons, observed rates of lepton pairs with invariant
   masses smaller than that of a $b$ quark are frequently used by collider
   experiments to determine or verify the $b$ purity of data samples used
   to measure $b$-hadron properties or to calibrate efficiencies for
   detecting $b$ hadrons. The present study shows that there is at least a
   difficulty in modeling rates and kinematical properties of such lepton
   pairs.
%%%%%%%%%%%%%%%%%%%%%%%%%%%%%
 \section{Acknowledgments}
%%%%%%%%%%%%%%%%%%%%%%%%%%%%%
   We thank the Fermilab staff, the CDF collaboration, and the technical
   staff of its participating Institutions for their contributions. This 
   work was supported by the U.S.~Department of Energy and the Istituto
   Nazionale di Fisica Nucleare. We warmly acknowledge the contribution
   of M.~Mazzucato and P.~Ronchese in comparing our simulation to the 
   DELPHI simulation and data.
%%%%%%%%%%%%%%%%%%%%%%%%%%%%%%%


\begin{thebibliography}{99}
\bibitem{thrust}   The thrust axis $\vec{T}$ is found maximizing the ratio
                    $\frac{\sum_{a} |\vec{p}_a \cdot \vec{T}|} 
                    {\sum_{a} |\vec{p}_a|}$ where $\vec{p}_a$ is the momentum
                    of each particle $a$ in the final state and $|\vec{T}|$=1.
\end{thebibliography}
 \end{document}